\documentclass[a4paper,11pt]{article}
\def \ka{\kappa_1}
\def \kb{\kappa_2}
\newcommand{\pard}[2]{\frac{\partial #1}{\partial #2}}  
\def \ccomma{\raise 2pt\hbox{,}}

\title{General solution for Hamiltonians with extended cubic and quartic 
potentials
\footnote{ \textit{Theoretical and Mathematical Physics},
to appear.
S2002/094. nlin.SI/0301011}
}
\author{C.~Verhoeven, M.~Musette and R.~Conte}

\begin{document}
\maketitle


\begin{abstract}
We integrate with hyperelliptic functions a two-particle Hamiltonian 
with quartic potential and additionnal linear and nonpolynomial terms 
in the Liouville integrable cases $1:6:1$ and $1:6:8$.
\end{abstract}

\section{Introduction}

The generalized H\'enon-Heiles Hamiltonian
\begin{equation}\label{HH:gham}
H=\frac{1}{2}(P_X^2+P_Y^2+c_1X^2+c_2Y^2)+aXY^2-\frac{b}{3}X^3+\mu Y^{-2},\quad 
\mu\hbox{ arbitrary}
\end{equation}
is known to be Liouville integrable  \cite{BSV:1982,CTW:1982,GDP:1982} for 
three sets of values of $(b/a,c_1,c_2)$ and to be related \cite{For:1991} 
to the stationary reduction of the following fifth order 
soliton equations: KdV$_5$, Sawada-Kotera
(SK) and Kaup-Kupershmidt (KK).

The canonical transformation \cite{Bak:1995} between the SK and KK cases for 
$\mu\neq 0$ allowed us \cite{VMC:2001} to define
the separating variables of the Hamilton-Jacobi equations and to derive the 
general solution of the equations of motion with
hyperelliptic functions.

The two-particle Hamiltonian with quartic potential
\begin{equation}\label{qua:gham}
H=\frac{1}{2}(P_X^2+P_Y^2)-\frac{1}{2}(aX^2+bY^2)+CX^4+BX^2Y^2+AY^4,
\end{equation}
is known to be Liouville integrable in four cases 
\cite{GDR:1983,Hie:1984,Hie:1987}:  
\begin{enumerate}
\item $A:B:C=1:2:1,a,b$ arbitrary, 
\item $A:B:C=1:6:1,a=b$, 
\item $A:B:C=1:6:8,a=4b$, 
\item $A:B:C=1:12:16,a=4b$.
\end{enumerate}

The extension of (\ref{qua:gham}) to include some linear and non polynomial 
terms which preserve
the Liouville integrability \cite{GDR:1984,Hie:1987} and confirm their 
connection with some
integrable soliton equations has been considered by various authors 
\cite{RRG:1994,Rom:1995}. 

The case $1$ with extra inverse square terms is equivalent to the 
travelling wave reduction of the Manakov system
\cite{Man:1973} of two coupled NLS equations. 
It corresponds to the particular case for two
particles of the extended Garnier system \cite{Gar:1919} which has been 
integrated by Wojciekowski \cite{Woj:1985}. 
The case $4$ can be associated \cite{Bak:1995} with a coupled 
KdV system
possessing a fifth order Lax pair
\cite{DrS:1981}.

The authors of \cite{BEF:1995,Bak:1995} have showed how cases $2$ and $3$
with some extra linear and inverse square terms 
are equivalent to the stationary reduction $\xi=x-ct$ of 
the Hirota-Satsuma system 
\cite{SaH:1982}:
\begin{equation}\label{qua:HSs}
r_t=\frac{1}{2}r_{xxx}+3rr_x-6ss_x,\quad s_t=-s_{xxx}-3rs_x
\end{equation}
and an other coupled KdV system
\begin{equation}\label{qua:cKdV} 
\left\{\begin{array}{ll}
f_t=&-\frac{1}{2}(2f_{xxx}+3ff_{xx}+3f_x^2-3f^2f_x+6fg_x+6gf_x)\ccomma\\[4pt]
g_t=&\frac{1}{4}(2g_{xxx}+12gg_x+6fg_{xx}+12gf_{xx}+18f_xg_x-6f^2g_x\\
    &+3f_{xxxx}+3ff_{xxx}+18f_xf_{xx}-6f^2f_{xx}-6ff_x^2)\ccomma
\end{array}
\right.
\end{equation}
In the present paper, using the canonical transformation \cite{Bak:1995} 
between these two extended cases, 
we show how the method we followed for the SK and KK cases of 
(\ref{HH:gham}) to define the
separating variables of the Hamilton-Jacobi equation can be applied here. Then 
the equations
of motion associated with  cases $2$ and $3$ of (\ref{qua:gham}) with extra 
linear and inverse
square terms are also integrated with hyperelliptic functions. 

In section \ref{section2}, we consider the SK and KK integrable cases of the 
H\'enon-Heiles Hamiltonian (\ref{HH:gham}) for $\mu\neq 0$ to explain 
the method we applied in \cite{VMC:2001} for integrating their equations of 
motion. 
In section \ref{section3} we consider the Hamiltonian (\ref{qua:gham}) with 
extended quartic potential in case $2$ and $3$ and recall the canonical
transformation \cite{Bak:1995}, which relates the case $1:6:1$ of 
(\ref{qua:gham}) (with extra terms of the form $\alpha X^{-2}+\beta Y^{-2}$) 
and the case $1:6:8$ (with extra terms like $\gamma X+\delta Y^{-2}$).

To the best of our knowledge,
the results of sections \ref{section4} and \ref{section5} are new. 
In section \ref{section4},
we explicitly integrate the equations of motion of the case $1:6:8$ for
$\delta\neq 0, \gamma=0$. 
In section \ref{section5} we integrate the $1:6:1$ case with
$\alpha=\beta\neq 0$ and we use the canonical transformation to transport those 
solutions to the
$1:6:8$ case for $\delta=0,\gamma\neq 0$.

\section{Integration of the cubic H\'enon-Heiles Hamiltonian}
\label{section2}

Let us consider the SK and KK H\'enon-Heiles Hamiltonians with their second 
integral, denoted as
$K_2^2$ and $k_2^2$, and their equations of motion \cite{GDR:1984,Hie:1987}:  
\begin{eqnarray}
\hbox{SK:}
&&\frac{b}{a}=-1,\quad, c_1=c_2,\quad a=\frac{1}{2},\quad U=X+c_2,\quad 
V=Y,\quad c=c_1c_2\\
\label{HH:HSK}
&&H\equiv K_1=\frac{1}{2}(P_U^2+P_V^2)
+\frac{1}{2}UV^2+\frac{1}{6}U^3-\frac{c}{2}U+\frac{\mu}{8V^2}\ccomma\\
\label{HH:KSK}
&&K_2^2=K_{2,0}^2+\frac{2}{3}\mu U+\mu\frac{P_U^2}{V^2}\ccomma\\
\label{HH:K0SK}
&&K_{2,0}=-2P_UP_V-U^2V-\frac{V^3}{3}+c V \ccomma\\
\label{HH:e2SK}
&&U''=-\frac{1}{2}(V^2+U^2)-\frac{c}{2}\ccomma\qquad 
V''=-UV+\frac{\mu}{4V^3}\ccomma
\\[10pt]
\hbox{KK:}
&&\frac{b}{a}=-16,\quad c_1=16c_2,\quad a=\frac{1}{4},\quad c=c_1c_2,\quad 
u=X+2c_2,\quad V=Y\\
\label{HH:HKK}
&&H\equiv k_1=\frac{1}{2}(p_u^2+p_v^2)+\frac{1}{4}uv^2+\frac{4}{3}u^3-cu+
\frac{1}{2}\frac{\mu}{v^2}\ccomma\\
\label{HH:KKK}
&&k_2^2=k_{2,0}^2+\frac{\mu}{3}u+2\mu\frac{p_v^2}{v^2}+\frac{\mu^2}{v^4}
\ccomma\\
\label{HH:K0KK}
&&k_{2,0}^2=p_v^4-\frac{1}{72}v^6-\frac{1}{12}u^2v^4+up_v^2v^2
-\frac{1}{3}p_up_vv^3+\frac{c}{12}v^4,\\
\label{HH:e2KK}
&&u''=-\frac{1}{4}v^2-4u^2+c,\qquad 
v''=-\frac{1}{2}uv+\frac{\mu}{v^3}\ccomma
\end{eqnarray}
which are equivalent to the stationary reduction $\xi=x-ct$ of the two 
integrable PDE's:
{\setlength\arraycolsep{1pt} \begin{eqnarray}
\hbox{SK: }&&U_t+(U_{xxxx}+5UU_{xx}+\frac{5}{3}U^3)_x=0,\\
\hbox{KK: }&&u_t+(u_{xxxx}+10uu_{xx}+\frac{20}{3}u^3+30u_x^2)_x=0.
\end{eqnarray}}
Both equations are connected to another fifth order integrable PDE 
\cite{FoG:1980}:
\begin{equation}\label{HH:fge}
w_t+(w_{4x}-5w_xw_{xx}-5w^2w_{xx}-5ww_x^2+w^5)_x=0.
\end{equation}
by the Miura transformation 
\begin{equation}
U=w_x-w^2,\qquad u=-w_x-\frac{1}{2}w^2.
\end{equation}
Solving the stationary reduction of (\ref{HH:fge}) for $w$ in function of 
$V,V'$ and $v,v'$ and defining the following expressions: 
{\setlength\arraycolsep{2pt}
\begin{eqnarray}
\lambda^2&=&-\mu,\
\Gamma=6(VK_{2,0}+\lambda P_U),\nonumber\\[6pt]
\Omega&=&
48(3v^4k_{2,0}^2+6\lambda uv^5p_v+12\lambda p_v^3v^3-\lambda v^6p_u
\nonumber\\
&&
+3\lambda ^2uv^4+18\lambda ^2v^2p_v^2+12\lambda^3vp_v+3\lambda^4),
\end{eqnarray}}
one obtains the following canonical transformation \cite{Bak:1995,BRW:1994}:
{\setlength\arraycolsep{2pt}
\begin{eqnarray}
\label{HHu:KKSK}
u
&=&
-\frac{3}{2}\left(-\frac{P_V}{V}+\frac{\lambda}{2V^2}\right)^2-U,\quad 
v=\frac{\sqrt{\Gamma}}{V}\ccomma
\\
\label{HHpu:KKSK}
p_u
&=&
\frac{1}{V^3}(3P_V^3+3UV^2P_V-P_UV^3)
\nonumber
\\
& &
 -\frac{3\lambda}{2V^6} 
\left(UV^4+3V^2P_V^2-\frac{3}{2}\lambda V P_V+\frac{\lambda^2}{4}\right),
\nonumber 
\\
p_v
&=&
\frac{1}{4V^2} \left(-2P_V+\frac{\lambda}{V}\right) \sqrt{\Gamma}
-\lambda \frac{V}{\sqrt{\Gamma}}\ccomma
\\
\label{HHU:SKKK}
U
&=&
-6\left(\frac{p_v}{v}+\frac{\lambda}{v^2}\right)^2-u,\quad 
V=\frac{\sqrt{\Omega}}{2v^4},
\\
\label{HHPU:SKKK}
P_U
&=&
\frac{1}{v^3}(12p_v^3+6uv^2p_v-v^3p_u)
\nonumber
\\
& &
+\frac{3\lambda}{v^6} (2uv^4+12v^2p_v^2+12\lambda v p_v+4\lambda^2)\ccomma
\\
P_V
&=&
-\frac{1}{v^5}\left(p_v+\frac{\lambda}{v}\right) \sqrt{\Omega}
+\lambda\frac{v^4}{\sqrt{\Omega}}\cdot
\end{eqnarray}
}
For $\mu=0$, we introduce the expressions (\ref{HHU:SKKK})--(\ref{HHPU:SKKK}) 
for $U,V,P_U,P_V$ in
the variables which separate the SK Hamiltonian
\begin{equation}\label{HHSK:sepv}
\begin{array}{ll}
Q_1=U+V,     &P_1=\frac{1}{2}(P_U+P_V)\ccomma\\[4pt]
Q_2=U-V,     &P_2=\frac{1}{2}(P_U-P_V)\cdot
\end{array}
\end{equation}
This defines the change of variables \cite{RGC:1993}
\begin{equation}\label{HHKK:sepv}
{\setlength\arraycolsep{10pt}\begin{array}{ll}
q_1=\displaystyle{-6\frac{p_v^2-k_{2,0}}{v^2}-u},
&p_1=\displaystyle{\frac{1}{2v^3}(12p_v^3+6uv^2p_v-v^3p_u-12p_vk_{2,0})}
\ccomma\\[6pt]
q_2=\displaystyle{-6\frac{p_v^2+k_{2,0}}{v^2}-u},
&p_2=\displaystyle{\frac{1}{2v^3}(12p_v^3+6uv^2p_v-v^3p_u+12p_vk_{2,0})}\ccomma
\end{array}}
\end{equation} 
that we apply on the KK Hamiltonian (\ref{HH:HKK}), taking account that for 
$\mu\neq 0$, $k_{2,0}$
is no more a constant of motion. Therefore, one has 
{\setlength\arraycolsep{2pt}\begin{eqnarray}
\label{HHKK:Hsep}
H&\equiv &k_1=
p_1^2+p_2^2+\frac{1}{12}(q_1^3+q_2^3)-\frac{c}{4}(q_1+q_2)+
\frac{\mu}{24}\frac{q_1-q_2}{k_{2,0}}\ccomma\\[6pt]
\label{HHKK:k0sep}
k_{2,0}&=&2(p_2^2-p_1^2)+\frac{1}{6}(q_2^3-q_1^3)-\frac{c}{2}(q_2-q_1),\\[6pt]
\label{HHKK:e1sep}
q_1'&=&2p_1+\frac{\mu}{6}\frac{(q_1-q_2)p_1}{k_{2,0}^2}\ccomma\\[6pt]
\label{HHKK:e2sep}
q_2'&=&2p_2-\frac{\mu}{6}\frac{(q_1-q_2)p_2}{k_{2,0}^2}.
\end{eqnarray}}
In this new setting of coordinates, it is important to note that,
introducing the expression
\begin{equation}
f(q_i,p_i)=2p_i^2+\frac{1}{6}q_i^3-\frac{c}{2}q_i \hbox{  for } i=1,2,
\end{equation} 
in (\ref{HHKK:Hsep}) and (\ref{HHKK:k0sep}), the Hamilton-Jacobi equation is 
separated for
$\mu$ arbitrary.

In the variables (\ref{HHKK:sepv}), the second invariant $k_2^2$ can be 
written in two equivalent ways:
\begin{eqnarray}
\label{HHKK:ksep1}
&&k_2^2=-\frac{\mu}{3}q_1
+\left(k_{2,0}+\frac{\mu}{12}\frac{q_1-q_2}{k_{2,0}}\right)^2,\\
\label{HHKK:ksep2}
\hbox{or}
&&k_2^2=-\frac{\mu}{3}q_2
+\left(k_{2,0}-\frac{\mu}{12}\frac{q_1-q_2}{k_{2,0}}\right)^2,
\end{eqnarray}
which, after the elimination of $\mu (q_1-q_2)/k_{2,0}$ between
(\ref{HHKK:Hsep})--(\ref{HHKK:ksep1}) and  
(\ref{HHKK:Hsep})--(\ref{HHKK:ksep2}), become
\begin{eqnarray}
\label{HHKK:ksep3}
&&k_2^2=-\frac{\mu}{3}q_1+\left(-4p_1^2-\frac{q_1^3}{3}+cq_1+2k_1\right)^2,\\
\label{HHKK:ksep4}
\hbox{or}
&&k_2^2=-\frac{\mu}{3}q_2+\left(4p_2^2+\frac{q_2^3}{3}-cq_2-2k_1\right)^2\cdot
\end{eqnarray}  
Then we can eliminate $p_1$ between (\ref{HHKK:ksep3}) and (\ref{HHKK:e1sep}),
and $p_2$ between (\ref{HHKK:ksep4}) and (\ref{HHKK:e2sep}) to obtain 
{\small
\begin{eqnarray}
\label{HHKK:e3sep}&&\hspace{-30pt}
q_1'=\sqrt{2k_1-\frac{q_1^3}{3}+cq_1-\sqrt{k_2^2+\frac{\mu}{3}q_1}}
\left(1+\frac{\mu}{3}\frac{q_1-q_2}{\left(\sqrt{k_2^2+\frac{\mu}{3}q_2}+
\sqrt{k_2^2+\frac{\mu}{3}q_1}\right)^2}\right)\ccomma\\
\label{HHKK:e4sep}&&\hspace{-30pt}
q_2'=\sqrt{2k_1-\frac{q_2^3}{3}+cq_2+\sqrt{k_2^2+\frac{\mu}{3}q_2}}\left(1-
\frac{\mu}{3}\frac{q_1-q_2}{\left(\sqrt{k_2^2+\frac{\mu}{3}q_2}+
\sqrt{k_2^2+\frac{\mu}{3}q_1}\right)^2}\right)\cdot
\end{eqnarray}}                       
For $\mu\neq 0$, setting:
\begin{equation}
s_1=\sqrt{3\frac{k_2^2}{\mu}+q_1},\qquad s_2=-\sqrt{3\frac{k_2^2}{\mu}+q_2},
\end{equation}
and defining
\begin{equation}\label{HH:hypc}
P(s)=2k_1-\frac{1}{3}\left(s^2-3\frac{k_2^2}{\mu}\right)^3
    +c\left(s^2-3\frac{k_2^2}{\mu}\right)-\sqrt{\frac{\mu}{3}}s,
\end{equation}
the equations (\ref{HHKK:e3sep})--(\ref{HHKK:e4sep}) become:
\begin{eqnarray}
&&
s_1'= \frac{\sqrt{P(s_1)}}{s_1-s_2}\ccomma\
s_2'=-\frac{\sqrt{P(s_2)}}{s_1-s_2}\cdot
\end{eqnarray}
They can be solved by inverting the hyperelliptic integrals
\begin{eqnarray}
\label{hyp1:HH}&&\int_{\infty}^{s_1}{\frac{\hbox{d}s}{\sqrt{P(s)}}}+
\int_{\infty}^{s_2}{\frac{\hbox{d}s}{\sqrt{P(s)}}}=k_3,\\
\label{hyp2:HH}&&\int_{\infty}^{s_1}{\frac{s\hbox{d}s}{\sqrt{P(s)}}}+
\int_{\infty}^{s_2}{\frac{s\hbox{d}s}{\sqrt{P(s)}}}=\xi+k_4.
\end{eqnarray}
and the general solution for the KK system is
\begin{equation}
\label{HH:sKK}
u=-\frac{1}{2}(s_1^2+s_2^2)+\frac{3}{\mu}k_2^2
  -\frac{3}{2}\left(\frac{s_1'+s_2'}{s_1+s_2}\right)^2,\qquad
v^2=\frac{2\sqrt{3\mu}}{s_1+s_2}\cdot
\end{equation}
For $\mu=0$ from (\ref{HHKK:e3sep})--(\ref{HHKK:e4sep}) and (\ref{HHKK:sepv}), 
one easily recovers the known solution \cite{RGC:1993} 
expressed with Weierstrass elliptic functions:
\begin{equation}
\label{HH:sKK0}
u=6(\wp_1+\wp_2)-\frac{3}{2}
\left(\frac{\wp_1'-\wp_2'}{\wp_1-\wp_2}\right)^2,\qquad
v^2=\displaystyle{\frac{k_{2,0}}{\wp_2-\wp_1}}\ccomma
\end{equation}
where $\wp_i$ satisfies the equation
\begin{eqnarray}
& & \wp_i'^2=4\wp_i^3-g_2\wp_i-g_3^{(i)},\quad i=1,2\\
\nonumber
& & g_2=\frac{c}{12}, \qquad 
    g_3^{(1)}=-\frac{1}{144}(2k_{1,0}-k_{2,0}),\quad
    g_3^{(2)}=-\frac{1}{144}(2k_{1,0}+k_{2,0}).
\end{eqnarray}
In the SK case, from (\ref{HH:sKK}) and the canonical transformation 
(\ref{HHU:SKKK}),  the general solution writes
\begin{equation}\label{HH:sSK}\begin{array}{ll}
U=\sqrt{-3}(s_1'+s_2')+s_1^2+s_1s_2+s_2^2-
\frac{3}{\mu}K_2^2,\\
V^2=-2\sqrt{-3}(s_1+s_2)(s_1s_1'+s_2s_2')
+2(s_1+s_2)^2\left(s_1^2+s_2^2-\frac{9K_2^2}{2\mu}\right)\ccomma
\end{array}
\end{equation}
which, in the particular case $\mu=0$, is merely
\begin{equation}\label{HH:sSK0}
U=-6\left(\wp_1+\wp_2\right),\qquad
V=-6\left(\wp_1-\wp_2\right)\cdot
\end{equation}

\section{Canonical transformation between extended $1:6:1$ and $1:6:8$
Hamiltonians}
\label{section3}

Let us consider the quartic Hamiltonian (\ref{qua:gham}) with extra linear and 
inverse quadratic terms in the
two following cases \cite{Hie:1987}:

{\setlength\arraycolsep{2pt} \begin{eqnarray}
1:6:1
\label{qua:H161}
&&H\equiv K_1=\frac{1}{2}(P_U^2+P_V^2)-\frac{1}{32}(U^4+6U^2V^2+V^4)\nonumber\\
&&\hspace{50pt}-\frac{c}{2}(U^2+V^2)+\frac{1}{2}(\frac{\alpha}{U^2}+\frac{\beta}
{V^2})\ccomma\\[5pt]
\label{qua:K161}
&&K_2^2=K_{2,0}^2-\beta U^2-\alpha V^2+4(\alpha\frac{P_V^2}{U^2}+
\beta\frac{P_U^2}{V^2})+\frac{4\alpha\beta}{U^2V^2}-4c(\alpha+\beta)
\ccomma\\[5pt] 
\label{qua:K0161}
&&K_{2,0}=2P_UP_V-\frac{1}{4}UV(U^2+V^2+8c)\ccomma\\[5pt] 
\label{qua:e1161}
&&U'=P_U,\qquad V'=P_V\ccomma\\
\label{qua:e2161}
&&U''=\frac{1}{8}U^3+\frac{3}{8}UV^2+cU+\frac{\alpha}{U^3}\quad
V''=\frac{1}{8}V^3+\frac{3}{8}U^2V+cV+\frac{\beta}{V^3}\ccomma\\[10pt]
1:6:8 
\label{qua:H168}
&&H\equiv k_1=\frac{1}{2}(p_u^2+p_v^2)-\frac{1}{16}(8u^4+6u^2v^2+v^4)\nonumber\\
&&\hspace{50pt}-\frac{c}{2}(4u^2+v^2)-\gamma u+\frac{\delta}{2v^2}\ccomma\\[5pt]
\label{quak168}
&&\begin{array}{ll}
k_2^2=&k_{2,0}^2+\frac{\delta^2}{v^4}-\gamma
v^2u^3-\frac{\gamma}{2}uv^4-\frac{\delta}{2}u^2-2\gamma^2v^2+2\delta
\frac{p_v^2}{v^2}-\frac{\delta}{4}v^2\\[3pt]
      &+4\gamma (vp_up_v-up_v^2)-4\gamma\delta\frac{u}{v^2}-2\delta c-4c\gamma 
v^2u\ccomma
\end{array}\\[5pt]
\label{qua:k0168}
&&\begin{array}{ll}
k_{2,0}^2=&(-p_v^2+\frac{1}{8}v^4+\frac{1}{4}u^2v^2-uvp_v+\frac{1}{2}p_uv^2+cv^2
)\\[3pt]
&\times(-p_v^2+\frac{1}{8}v^4+\frac{1}{4}u^2v^2+uvp_v-\frac{1}{2}p_uv^2+cv^2)
\ccomma
\end{array}\\[5pt]
\label{qua:e1168}
&&u'=p_u,\qquad v'=p_v\ccomma\\[5pt]
\label{qua:e2168}
&&u''=2u^3+\frac{3}{4}uv^2+4cu+\gamma,\quad
v''=\frac{1}{4}v^3+\frac{3}{4}u^2v+cv+\frac{\delta}{v^3}\ccomma 
\end{eqnarray}}
where $K_1,K_2$ and $k_1,k_2$ are the constants of motion of both systems. 

The equations of motion are respectively equivalent to the stationary 
reduction $\xi=x-ct$ of the PDE's (\ref{qua:HSs}) and
(\ref{qua:cKdV}) with the correspondence:
\begin{eqnarray}
\label{qua:tr161}&&r=-\frac{1}{4}(U^2+V^2+4c),\qquad 
s=\frac{1}{8}(U^2-V^2)\ccomma\\
\label{qua:tr168}&&f=u,\qquad g=-\frac{1}{4}(v^2+2u^2+2p_u+4c)\cdot
\end{eqnarray}
Both PDE's (\ref{qua:HSs}) and (\ref{qua:cKdV}) possess a fourth order Lax pair 
with scattering
operators $L$ and $\widetilde{L}$ which factorize as follows
\cite{Bak:1995,BEF:1995},
\begin{eqnarray}
\label{qua:L161}
&&\begin{array}{ll} 1:6:1\hspace{7pt}
L&=(\partial_x^2+r+s)(\partial_x^2+r-s)\\
 &=(\partial_x-v_1)(\partial_x+v_1)(\partial_x+v_2)(\partial_x-v_2)\ccomma
\end{array}\\
\label{qua:L168}
&&\begin{array}{ll}1:6:8\hspace{7pt}
\widetilde{L}&=(\partial_x^2+f\partial_x+f_x+g)(\partial_x-f\partial_x+g)\\[3pt]
 &=(\partial_x+v_1)(\partial_x+v_2)(\partial_x-v_2)(\partial_x-v_1)\ccomma
\end{array}
\end{eqnarray}
and yield the Miura maps:
\begin{eqnarray}
\label{qua:Miu161}
&&1:6:1\hspace{7pt}r=\frac{1}{2}(v_{1,x}-v_{2,x}-v_1^2-v_2^2),\quad 
s=\frac{1}{2}(v_{1,x}+v_{2,x}-v_1^2+v_2^2)\ccomma\\
\label{qua:Miu168}
&&1:6:8\hspace{7pt}f=v_1+v_2,\qquad g=v_1v_2-v_{1,x}\ccomma
\end{eqnarray}
while $v_1$ and $v_2$ are solutions of the system of PDE's
\begin{equation}\label{qua:emod}
\begin{array}{ll}
v_{1,t}=\frac{1}{8}(-2v_{1,xx}-6v_{2,xx}-12v_2v_{2,x}-12v_1v_{2,x}-12v_1v_2^2+4v
_1^3)_x\ccomma\\[5pt]
v_{2,t}=\frac{1}{8}(-2v_{2,xx}-6v_{1,xx}+12v_1v_{1,x}+12v_2v_{1,x}-12v_2v_1^2+4v
_2^3)_x.
\end{array}
\end{equation}

{}From the relations (\ref{qua:Miu161})--(\ref{qua:Miu168}), 
the transformations (\ref{qua:tr161})--(\ref{qua:tr168}) and the
equations of motion (\ref{qua:e1161})--(\ref{qua:e2161}), 
(\ref{qua:e1168})--(\ref{qua:e2168}), the stationary reduction of
(\ref{qua:emod}) can be solved for $v_1$ and $v_2$ in function of the canonical 
variables of the two extended systems:
\begin{eqnarray}
1:6:1&&v_1=-\frac{P_U}{U}+\frac{\sqrt{-\alpha}}{U^2},\qquad 
v_2=\frac{P_V}{V}+\frac{\sqrt{-\beta}}{V^2}\ccomma\\
1:6:8&&v_1=\frac{u}{2}+\frac{p_v}{v}+\frac{\sqrt{-\delta}}{v^2},\qquad
v_2=\frac{u}{2}-\frac{p_v}{v}-\frac{\sqrt{-\delta}}{v^2}\ccomma
\end{eqnarray}
such that, defining 
\begin{eqnarray}
\label{qua:dect}
&&\alpha=-\kappa_1^2,\quad \beta=-\kappa_2^2,\quad 
\delta=-(\kappa_2-\kappa_1)^2,\quad
\gamma=\frac{1}{2}(\kappa_1+\kappa_2)\ccomma\\
&&
\Omega=-\frac{1}{4}(U^2+V^2)+\frac{2}{UV}(P_UP_V+\kappa_2\frac{P_U}{V}
-\kappa_1\frac{P_V}{U}-\frac{\kappa_1\kappa_2}{UV})-2c\ccomma\\
&&
\Gamma_{\mp}=-4\Big(\mp 2p_u+\frac{1}{2}v^2+u^2-4\frac{p_v^2}{v^2}
\pm 4\frac{up_v}{v}+8(\kappa_2-\kappa_1)\frac{p_v}{v^3}
\nonumber
\\
&&
\phantom{\Gamma_{\mp}=-4\Big(}
\mp 4(\kappa_2-\kappa_1)\frac{u}{v^2}-4\frac{(\kappa_2-\kappa_1)^2}{v^4}+
4c\Big)\ccomma
\end{eqnarray}
the canonical transformation between the $1:6:1$ and $1:6:8$ cases becomes 
\cite{Bak:1995}
\begin{eqnarray}
&&u=-\frac{P_U}{U}+\frac{P_V}{V}+\frac{\kappa_1}{U^2}+\frac{\kappa_2}{V^2}
\ccomma\nonumber\\
\label{qua:168161u}
&&p_u=-\frac{1}{4}(V^2-U^2)+\frac{P_U^2}{U^2}-\frac{P_V^2}{V^2}
-2\kappa_1\frac{P_U}{U^3}
-2\kappa_2\frac{P_V}{V^3}+\frac{\kappa_1^2}{U^4}-\frac{\kappa_2^2}{V^4}\ccomma\\
[5pt]
\label{qua:168161v}
&&v=2\sqrt{\Omega}\qquad 
p_v=\sqrt{\Omega}\left(-\frac{P_U}{U}-\frac{P_V}{V}+\frac{\kappa_1}{U^2}
-\frac{\kappa_2}{V^2}\right)-\frac{\kappa_2-\kappa_1}{2\sqrt{\Omega}}
\ccomma\\[5pt]
\label{qua:161168U}
&&U=\frac{1}{2}\sqrt{\Gamma_-},\qquad
P_U=-\frac{1}{2}\sqrt{\Gamma_-}\left(\frac{u}{2}+\frac{p_v}{v}
-\frac{\kappa_2-\kappa_1}{v^2}\right)+2\kappa_1\sqrt{\Gamma_-}\ccomma\\[5pt]
\label{qua:161168V}
&&V=\frac{1}{2}\sqrt{\Gamma_+},\qquad
P_V=\frac{1}{2}\sqrt{\Gamma_+}\left(\frac{u}{2}-\frac{p_v}{v}
+\frac{\kappa_2-\kappa_1}{v^2}\right)-2\kappa_2\sqrt{\Gamma_+}\cdot
\end{eqnarray}

\section{General solution of the $1:6:8$ Hamiltonian for $\gamma=0$}
\label{section4}

For $\alpha=\beta=0$, one introduces the transformation 
(\ref{qua:161168U})--(\ref{qua:161168V}) in the variables
\begin{equation}
{\setlength\arraycolsep{6pt}\begin{array}{ll}\label{q1610:sepv}
Q_1=\frac{1}{2}(U+V)^2, 
&P_1=\displaystyle{\frac{1}{2}\frac{P_U+P_V}{U+V}}\ccomma\\[4pt]
Q_2=\frac{1}{2}(U-V)^2, 
&P_2=\displaystyle{\frac{1}{2}\frac{P_U-P_V}{U-V}}\ccomma
\end{array}}
\end{equation}
which separated the $1:6:1$ Hamiltonian (\ref{qua:H161}) in the polynomial 
case. 
This defines the following change of variables:
\begin{eqnarray}
&&q_1=4\frac{k_{2,0}+p_v^2}{v^2}-\frac{v^2}{2}-u^2-4c\ccomma\\
&&p_1=\frac{-8vp_vp_u+8up_v^2+2v^2u^3+u(v^4+8k_{2,0})+8cuv^2}{16v(vp_u-2up_v)},
\\
&&q_2=4\frac{-k_{2,0}+p_v^2}{v^2}-\frac{v^2}{2}-u^2-4c,\\
&&p_2=\frac{-8vp_vp_u+8up_v^2+2v^2u^3+u(v^4-8k_{2,0})+8cuv^2}{16v(vp_u-2up_v)}
\ccomma
\end{eqnarray}

that we apply on the Hamiltonian system (\ref{qua:H168}) for $\ka\kb\neq 0$, 
i.e. when $k_{2,0}$ is no more a constant of
motion. This transformation restricted to $\kb=-\ka$ ($\delta=-4\ka^2,\\
\gamma=0$) defines the coordinates which separate
\cite{RRG:1994,Rom:1995} the Hamilton-Jacobi equation. Indeed, one has
\begin{equation}
\label{qua168a:Hsep}
H\equiv k_1=\frac{1}{16}\Big(32q_1p_1^2-q_1^2+32q_2p_2^2-q_2^2-8c(q_1+q_2)\Big)-
\ka^2\frac{(q_1-q_2)}{4k_{2,0}}\ccomma
\end{equation}
with
\begin{eqnarray}
\label{qua168a:k0sep}
&&k_{2,0}=\frac{1}{8}\Big(32q_1p_1^2-q_1^2-32q_2p_2^2+q_2^2+8c(q_2-q_1)\Big)
\ccomma\\
\label{qua168a:e1sep}
&&q_1'=4q_1p_1+2\ka^2\frac{(q_1-q_2)q_1p_1}{k_{2,0}^2}\ccomma\\
\label{qua168a:e2sep}
&&q_2'=4q_2p_2-2\ka^2\frac{(q_1-q_2)q_2p_2}{k_{2,0}^2}\ccomma
\end{eqnarray}
such that defining 
\begin{displaymath}
f(q_i,p_i)\equiv 32q_ip_i^2-q_i^2-8cq_i,\quad i=1,2
\end{displaymath}
we have the following separated Hamilton-Jacobi equation:
\begin{equation}
f(q_1,p_1)^2-f(q_2,p_2)^2-32\ka^2(q_1-q_2)=16k_1\left(f(q_1,p_1)-f(q_2,p_2)
\right),\quad
p_i=\pard{S}{q_i}.
\end{equation}
In analogy with the Kaup-Kupershmidt case, we can write the second invariant in 
two equivalent ways:
\begin{eqnarray}
\label{qua168a:ksep1}
&&k_2^2=-2\ka^2q_1+\bigg(k_{2,0}+\frac{\ka^2(q_1-q_2)}{2k_{2,0}}\bigg)^2
\ccomma\\
\hbox{or}
\label{qua168a:ksep2}
&&k_2^2=-2\ka^2q_2+\bigg(k_{2,0}-\frac{\ka^2(q_1-q_2)}{2k_{2,0}}\bigg)^2\ccomma
\end{eqnarray}
which allow us to eliminate $\ka^2(q_1-q_2)/k_{2,0}$ between
(\ref{qua168a:Hsep})--(\ref{qua168a:ksep1}) and 
(\ref{qua168a:Hsep})--(\ref{qua168a:ksep2}). Next,
we eliminate $p_1$ and $p_2$ between those two resulting expressions and the 
equations
(\ref{qua168a:e1sep}), (\ref{qua168a:e2sep}) and obtain:
\small
\begin{eqnarray}
\label{qua168a:e3sep}
&&\hspace{-10pt}q_1'=q_1\sqrt{\frac{q_1}{2}+4c+4\frac{k_1}{q_1}+
\frac{2}{q_1}\sqrt{k_2^2+2\ka^2q_1}}\Bigg(1+\frac{2\ka^2(q_1-q_2)}
{\left(\sqrt{k_2^2+2\ka^2q_1}+\sqrt{k_2^2+2\ka^2q_2}\right)^2}\Bigg),\\[6pt]
\label{qua168a:e4sep}
&&\hspace{-10pt}q_2'=q_2\sqrt{\frac{q_2}{2}+4c+4\frac{k_1}{q_2}-
\frac{2}{q_2}\sqrt{k_2^2+2\ka^2q_2}}
\Bigg(1-\frac{2\ka^2(q_1-q_2)}{\left(\sqrt{k_2^2+2\ka^2q_1}
+\sqrt{k_2^2+2\ka^2q_2}\right)^2}\Bigg).
\end{eqnarray}
\normalsize
For $\ka=0$ the general solution of equations 
(\ref{qua168a:e3sep})--(\ref{qua168a:e4sep}) can be
expressed in terms of Weierstrass elliptic functions:
\begin{eqnarray}
&&q_1+\frac{8}{3}c
=8\wp\Big(\xi-\xi_1,\frac{4}{3}c^2-\frac{k_1}{2}-\frac{k_2}{4},
     \frac{c}{12}(2k_1+k_2-\frac{32}{9}c^2)\Big)
\equiv 8\wp_1\ccomma\\
&&q_2+\frac{8}{3}c
=8\wp\Big(\xi-\xi_2,\frac{4}{3}c^2-\frac{k_1}{2}+\frac{k_2}{4},
     \frac{c}{12}(2k_1-k_2-\frac{32}{9}c^2)\Big)
\equiv 8\wp_2\cdot
\end{eqnarray}
For $\ka\neq 0$ setting in (\ref{qua168a:e3sep})--(\ref{qua168a:e4sep}):
\begin{equation}
s_1=\sqrt{\frac{1}{2\ka^2}k_2^2+q_1}\hspace{7pt}\hbox{
and }\hspace{7pt}s_2=-\sqrt{\frac{1}{2\ka^2}k_2^2+q_2},
\end{equation}
and defining 
\begin{equation}
\widetilde{P}(s)\equiv 
\frac{1}{2}\left(s^2-\frac{k_2^2}{2\ka^2}\right)^3
+4c\left(s^2-\frac{k_2^2}{2\ka^2}\right)^2
+(4k_1+2\sqrt{2}\ka s)\left(s^2-\frac{k_2^2}{2\ka^2}\right)\ccomma
\end{equation}
we obtain the equations,
\begin{eqnarray}
&&\hspace{-15pt}
s_1'= \frac{\sqrt{\widetilde{P}(s_1)}}{s_1-s_2}\ccomma\
s_2'=-\frac{\sqrt{\widetilde{P}(s_2)}}{s_1-s_2}\ccomma\
\end{eqnarray}
which  are solved with the inversion of the hyperelliptic integrals
\begin{eqnarray}
\label{qua:hyp1}
&&\int_{\infty}^{s_1}{\frac{\hbox{d}s}{\sqrt{\widetilde{P}(s)}}}+
\int_{\infty}^{s_2}{\frac{\hbox{d}s}{\sqrt{\widetilde{P}(s)}}}=k_3,\\
\label{qua:hyp2}
&&\int_{\infty}^{s_1}{\frac{s\hbox{d}s}{\sqrt{\widetilde{P}(s)}}}+
\int_{\infty}^{s_2}{\frac{s\hbox{d}s}{\sqrt{\widetilde{P}(s)}}}=\xi+k_4\cdot
\end{eqnarray}
Therefore, for $\delta$ arbitrary, $\gamma=0$, the general solution of the 
$1:6:8$ Hamiltonian is
defined with symmetric combinations of $s_1,s_2$
\begin{equation}\label{qua168a:gs}
u^2=-\frac{1}{2}(s_1^2+s_2^2)+\frac{k_2^2}{2\ka^2}
+\left(\frac{s_1'+s_2'}{s_1+s_2}\right)^2-\frac{2\sqrt{2}\ka}{s_1+s_2}-4c,
\quad
v^2=\frac{4\sqrt{2}\ka}{s_1+s_2}\ccomma
\end{equation}
and is a single-valued function of the complex variable $\xi$. 
For $\ka=0$ it degenerates into
\begin{equation}\label{qua168:gs0}
u^2=-4(\wp_1+\wp_2)+\left(\frac{\wp_1'-\wp_2'}{\wp_1-\wp_2}\right)^2
+\frac{k_{2,0}}{2(\wp_2-\wp_1)}-\frac{4}{3}c,\qquad
v^2=\frac{k_{2,0}}{\wp_1-\wp_2}\cdot
\end{equation}

\section{General solution of the extended $1:6:1$ Hamiltonian}
\label{section5}
Now, we again use the canonical transformation 
(\ref{qua:161168U})--(\ref{qua:161168V}) to find the
general solution of the $1:6:1$ Hamiltonian.

{\setlength\arraycolsep{2pt}
For $\ka=\kb=0$, we introduce the expressions (\ref{qua168:gs0}) in 
(\ref{qua:161168U})--(\ref{qua:161168V}) and obtain
the solution of the equations of motion
\begin{equation}
U^2+V^2=8(\wp_1+\wp_2-\frac{2}{3}c),\qquad UV=4(\wp_1-\wp_2)\cdot
\end{equation}
Next, for $\alpha\beta\neq 0$, with the restriction $\alpha=\beta$, 
we obtain the expressions
\begin{eqnarray}
\label{qua161a:gs1}
U^2+V^2&=&-2\sqrt{2}(s_1'+s_2')+2(s_1^2+s_2^2+s_1s_2)-\frac{K_2^2}{\ka^2}
\ccomma\\
\label{qua161a:gs2}
U^2V^2&=&
-2\sqrt{2}(s_1+s_2)(s_1s_1'+s_2s_2'-2 \ka)
\nonumber
\\
&&+2 (s_1+s_2)^2 \left(s_1^2+s_2^2-
\frac{3}{4}\frac{K_2^2}{\ka^2}+4c\right)\ccomma
\end{eqnarray}
which depend on symmetric combinations of $s_1,s_2$ 
and therefore are single-valued functions of $\xi$.
}

Since the Hamiltonian (\ref{qua:H161}) for $\alpha=\beta$ is an even function 
of $\ka$, we apply on
(\ref{qua161a:gs1})--(\ref{qua161a:gs2}) the canonical transformation 
(\ref{qua:168161u})--(\ref{qua:168161v}) for $\ka=\kb$,
and obtain the general solution of the $1:6:8$ Hamiltonian in the extended case 
$\delta=0$.

\section{Conclusion}

Romeiras \cite{Rom:1995b} described a procedure which relates the extended 
cubic potential 
$\widetilde{V}_3(u,v)=V_3(u,v)+\mu v^{-2}/2$ in the KK case
with the extended quartic potential
$\widetilde{V}_4(u,v)=V_4(u,v)+\delta v^{-2}/2$
in the $1:6:8$ case.
This explains why the combination $u^2+v^2/2+4c$ of the solutions 
(\ref{qua168a:gs}) is identical to
the solution (\ref{HH:sKK}).
The question remains open to extrapolate the present results to the case
$\alpha\not=\beta$, i.e. $\gamma \delta\not=0$.

\section*{Acknowledgments}

CV, 
a research assistant of the Fund for Scientific Research, Flanders,
is grateful to the organisers for their financial support
to attend the conference.
The financial support of the Tournesol grant T99/040,
the IUAP Contract No.~P4/08 funded by the Belgian government,
and the CEA, is gratefully acknowledged.

\end{document}